\documentstyle[11pt,newpasp,twoside,epsf]{article}
\def\HST{{\it HST}}

\def\ASCA{{\it ASCA}}

\def\kms{\ifmmode {\rm km\ s}^{-1} \else km s$^{-1}$\fi}
\def\Msun{\ifmmode M_{\odot} \else $M_{\odot}$\fi}
\def\Lsun{\ifmmode L_{\odot} \else $L_{\odot}$\fi}

\def\gtsim{\raisebox{-.5ex}{$\;\stackrel{>}{\sim}\;$}}
\def\vFWHM{\ifmmode v_{\mbox{\tiny FWHM}} \else
            $v_{\mbox{\tiny FWHM}}$\fi}
\def\Halpha{\ifmmode {\rm H}\alpha \else H$\alpha$\fi}
\def\Hbeta{\ifmmode {\rm H}\beta \else H$\beta$\fi}
\def\Hgamma{\ifmmode {\rm H}\gamma \else H$\gamma$\fi}
\def\Hdelta{\ifmmode {\rm H}\delta \else H$\delta$\fi}
\def\Lya{\ifmmode {\rm Ly}\alpha \else Ly$\alpha$\fi}
\def\Lyb{\ifmmode {\rm Ly}\beta \else Ly$\beta$\fi}

\def\ciii{\ifmmode {\rm C}\,{\sc iii} \else C\,{\sc iii}\fi}
\def\civ{\ifmmode {\rm C}\,{\sc iv} \else C\,{\sc iv}\fi}

\def\nv{N\,{\sc v}}

\def\o5007{[O\,{\sc iii}]\,$\lambda5007$}

\def\ovi{O\,{\sc vi}}

\def\siIV{Si\,{\sc iv}}

\def\caii{Ca\,{\sc ii}}

\begin{document}

\title{The Structure and Energetics of Active Galactic Nuclei}
\author{Bradley M. Peterson}
\affil{Department of Astronomy, The Ohio State University,
140 West 18th Avenue, Columbus, OH 43210}
\begin{abstract}
The black-hole/accretion-disk paradigm for active galactic
nuclei (AGNs) is now reasonably secure, but there are still important
unresolved issues, some of which will require the
capabilities of an 8 to 10-m class UV/optical space-based telescope.
Imaging spectroscopy with a diffraction-limited large telescope will
be required to measure AGN black-hole masses from stellar dynamics for
direct comparison with reverberation mapping-based masses.  High
spectral resolution in the UV is required to determine the mass and
kinetic energy of the outflows observed in the absorption spectra of
AGNs and to understand the energetics of the accretion process. As
with ground-based astronomy, however, effective use of a large
UV/optical space telescope requires complementary smaller facility
instruments; a meter-class UV spectroscopic telescope, for example,
can fit into a Medium Explorer budget.
\end{abstract}

\section{Key Questions in AGN Astrophysics}
Not long after the discovery of quasars, it was realized that the
fundamental source of energy for these objects must in fact be
gravitation, and fairly straightforward arguments led to the
long-standing paradigm of a supermassive black hole (SMBH) surrounded
by an accretion disk. Observational evidence has been
rather ambiguous, although there were early strong clues,
such as the near-UV/optical ``big blue bump''
(Shields 1978; Malkan \& Sargent 1982) and the rapid X-ray variability,
that supported the model if only because they
defied other explanation. But it is only within the last several
years that the circumstantial evidence has accumulated to the point
that few doubters remain. While there is now general, though not
unanimous, agreement about the fundamental nature of AGNs,
we cannot claim any real understanding of the quasar phenomenon
until we successfully address a number of key questions,
including the following:
\begin{enumerate}
\item {\em What are the masses of AGN black holes?} As described
below, we are making progress, but there are
still areas where our understanding is dangerously incomplete,
especially with regard to the magnitude of possible systematic
errors.
\item {\em What are the energetics of the accretion process?}
In particular, for the various types of AGNs
what are the accretion rates and radiative efficiencies
and how do these scale with luminosity?
How much of the output is in the form of 
kinetic energy (e.g., jets and absorbing gas) as opposed to radiation?
\item {\em How does the AGN mass function evolve over time?}
Does the accretion process contribute significantly to
black-hole growth, and how do black-hole demographics evolve
with time? 
\item {\em What is the nature of the line-emitting and absorbing gas 
in AGNs?} There is good reason to believe that these 
are somehow related to the accretion process, but there is no
standard paradigm for the origin and role of these components
in AGNs. This is one of the remaining
outstanding mysteries in AGN structure.
\end{enumerate}

\section{Masses of Black Holes in AGNs}
Ironically enough, the first reasonably
secure SMBH masses were not measured in AGNs, but in quiescent
galaxies, in which stellar or nuclear gas dynamical methodologies were
employed (see Kormendy \& Richstone 1995). The first and still
highest precision AGN SMBH mass was determined from megamaser
motions in NGC 4258 (Miyoshi et al.\ 1995). It was the fortuitious
combination of source geometry and inclination that made this 
possible and, unfortunately, the method is not generally applicable 
to a broader range of AGNs. Most SMBH mass
determinations in AGNs are based on emission-line reverberation mapping
(Blandford \& McKee 1982; Peterson 1993, 2001), which at this
stage remains a rather crude tool; 
these masses are systematically uncertain to at least a factor of a few, and
more accurate determination of these masses is an
important current problem in AGN astrophysics.

\subsection{Reverberation-Based Black-Hole Masses}
Reverberation mapping makes use of the natural variability
of AGNs to probe the central structure. The broad emission
lines that dominate the UV/optical spectra of AGNs vary
in response to changes in the continuum flux with a time
delay, or lag, $\tau$ that reflects the light-travel time across
the broad-line region (BLR). If gravity dominates the
dynamics of the BLR, then by combining the size of the
BLR $c\tau$ (as measured in a particular emission line) with
the Doppler width $V$ of the line, we can make a virial
estimate of the central mass
\begin{equation}
M_{\rm BH} = \frac{k c\tau V^2}{G},
\end{equation}
where the constant $k$ depends on the structure, kinematics, and 
projection (inclination) of the line-emitting region. 
It is the factor $k$, and thus the level of possible systematic 
uncertainties in the reverberation method, that remains unknown.

While the fundamental assumption that the gravitational field
of the SMBH determines the BLR gas dynamics is unproven, there
are nevertheless good reasons to believe that the
reverberation-based black-hole masses are meaningful and 
that the systematic errors are not large enough to render
the method useless.  First, different emission
lines within an individual source give consistent virial mass
estimates, i.e., the emission-line time lags and line widths
show a virial relationship, $\tau \propto V^{-2}$. This has now been
demonstrated for four AGNs 
(Peterson \& Wandel 1999, 2000; Onken \& Peterson 2002), and
constitutes the best evidence to date that the BLR motions
are determined primarily by gravity. The relationship is 
surprisingly tight given that the line-emitting regions for
the high-ionization and low-ionization
lines are not necessarily expected to have similar geometries. Second,
AGNs follow the same relationship between
black-hole mass and stellar bulge velocity dispersion
that is seen in quiescent galaxies (Gebhardt et al.\ 2000b;
Ferrarese et al.\ 2001),
constituting a strong demonstration that reverberation-based
masses are comparable to stellar-dynamical masses
in accuracy.

There are two important things that need to be done to secure our
understanding of black-hole masses in AGNs and their relationship to
the black holes in quiescent galaxies:
\begin{enumerate}
\item More intensive reverberation-mapping experiments that will allow 
determination of
detailed two-dimensional transfer functions for multiple 
emission lines. This will  lead to an 
understanding of the geometry and kinematics of the BLR
and thus allow us 
to assess the magnitude of possible errors in reverberation-based 
masses from less well-sampled experiments. 
The data requirements for such experiments are 
well-understood (Horne et al.\ 2002)
and there is still much that can be done with smaller facilities.
\item Direct comparison of stellar-dynamical and reverberation-based 
masses by applying 
both methods to common AGNs. This is in fact a very challenging 
exercise that can be 
attacked with an 8-m class diffraction-limited optical telescope, as 
described below.
\end{enumerate}

\subsection{AGN Black Hole Masses from Stellar Dynamics}

A useful criterion for obtaining an accurate stellar-dynamical black-hole mass 
is spatially resolving the black-hole radius of influence
\begin{equation}
r_* = \frac{G M_{\rm BH}}{\sigma^2},
\end{equation}
where $\sigma$ is the host galaxy bulge velocity dispersion.
Especially for Type 1 AGNs, 
the best stellar absorption features for velocity dispersion
measurement are the \caii\ triplet lines around 8600\,\AA, in no small
part because of the favorably low contrast between the
AGN light and starlight in this spectral region. 
To resolve $r_*$ at wavelength $\lambda$, we require
\begin{equation}
{\rm FWHM} = \frac{1.22\lambda}{D} < \frac{r_*}{d},
\end{equation}
where $D$ is the telescope aperture and $d$ is the distance to the AGN. 
We can eliminate $\sigma$ from eq.\ (2) by using the
now well-known relationship between black-hole mass and bulge velocity 
dispersion (Ferrarese \& Merritt 2000; Gebhardt et al.\ 2000a).
The parameterization of Ferrarese (2002) gives
\begin{equation}
\frac{M_{\rm BH}}{\Msun} = 1.66 \times 10^8 \left(\frac{\sigma}
{200\,{\rm km\ s}^{-1}}\right)^{4.58},
\end{equation}
and by combining eqs.\ (2) and (4), we obtain the criterion
\begin{equation}
\label{eq:masscrit}
\frac{M_{\rm BH}}{\Msun} > 1.1 \times 10^6 \left[ \frac{d{\rm (Mpc)}}
{D{\rm (m)}} \right] ^{1.78}. 
\end{equation}
This is the smallest black-hole mass that can be measured by
stellar-dynamical methods as a function of distance $d$ and telescope 
diameter $D$.
The strong dependence on distance favors measurement of nearby 
lower-mass black-holes over higher-mass distant black holes.

\begin{figure}
\plotone{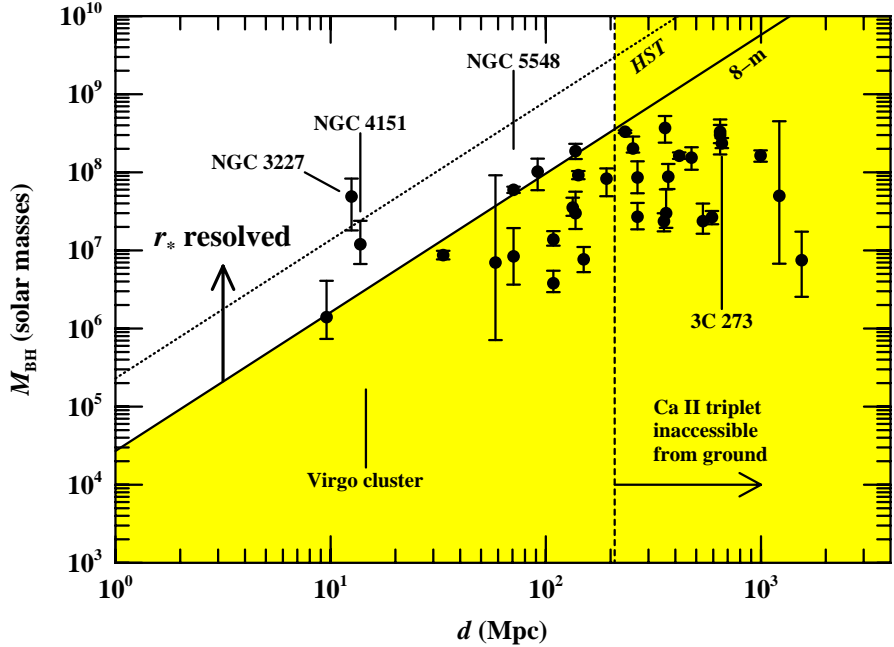}
\caption{The diagonal lines show the minimum detectable
black-hole mass as a function of distance for
\HST\ (dotted line) and an 8-m diffraction-limited telescope
(solid line), based on stellar-dynamical measurements of the
\caii\ triplet lines. At redshifts 
$z > 0.06$, the \caii\ triplet lines are inaccessible
from the ground due to water vapor in the Earth's atmosphere.
The individual points show AGNs for which reverberation-based
mass estimates are available.}
\end{figure}

We plot eq.\ (\ref{eq:masscrit}) in Figure 1 for two cases,
(a) for {\em Hubble Space
Telescope (HST)} with $D = 2.4$\,m, and (b) for an diffraction-limited
8-m telescope. SMBH masses above these respective lines are large enough to be
measurable by stellar-dynamical methods. Also plotted on this diagram
are masses and distances for reverberation-mapped AGNs.
This diagram shows that no more than two reverberation-mapped AGNs, 
NGC 3227 and NGC 4151, are even possible
candidates for measurement of their central masses by stellar-dynamical 
methods with \HST. On the other hand, with
a diffraction-limited 8-m telescope, a half-dozen or so
AGNs are amenable to stellar-dynamical measurements of their SMBH masses.
This is a marginally large enough sample to compare the stellar-dynamical
and reverberation-based masses in a definitive way 
and to allow a reasonable assessment of the currently unknown 
systematic errors in the reverberation-based masses.

We note that attempts have been made to observe both NGC~3227 
and NGC~4151, the two reverberation-mapped AGNs in which 
the black-hole radius of influence might be resolvable
with \HST\ (Gary Bower, PI). Neither attempt has been successful, in 
both cases because at the time the AGN was observed it was in an 
anomalously bright state, swamping the \caii\ triplet absorption lines
in the host-galaxy spectrum. Spatially resolved spectroscopy of both
of these sources should be re-attempted with \HST, but perhaps on a 
target-of-opportunity basis when the active nucleus is in a relatively 
faint state.

An obvious question to ask is why must these measurements be made from space,
since the \caii\ triplet is accessible from the ground? The main reason 
is that the strong contamination by the point-like nuclear source 
requires a truly diffraction-limited optical system. Ground-based 
adaptive optics systems are inadequate because of their low Strehl 
ratios (ratio of core intensity relative to a diffraction-limited 
point-spread function). Even in the low-luminosity AGNs, the weak 
stellar absorption features are completely washed out 
by the bright nuclear light.
A second reason to observe from space is that at a fairly low
redshift ($z \gtsim 0.06$),
the \caii\ triplet lines are redshifted into the strong telluric water 
vapor lines, thus making an already very difficult observation 
virtually impossible.

It is also worth noting that it is only Type 1 AGNs (i.e., those
with prominent broad lines in the UV/optical) that are amenable
to reverberation mass measurements; on the other hand, of course,
this is an important method that needs further development since
it can be applied to Type 1 AGNs at arbitrary distance.
Black-hole masses can be measured
by megamaser motions in only very special cases. Thus direct
measurement of the central masses of most local galaxies, including
many AGNs and related objects (e.g., Type 2 AGNs and LINERs),
will require stellar-dynamical studies. These, we see from
Fig.\ 1, are capable of measuring SMBH masses
for $M_{\rm BH} > 10^8\,\Msun$ out to a distance of 
$\sim100$\,Mpc. 

Comparison of stellar-dynamical and reverberation-based masses
is of critical importance. Once this has been effected, we will
be able to employ (a) reverberation-mapping methods to distant
AGNs and (b) secondary methods that are tied to reverberation
(e.g., Wandel, Peterson, \& Malkan 1999, Vestergaard 2002)
to estimate with confidence
masses of the black holes in distant quasars and thus
address how the quasar mass function evolves with time.

\section{X-Ray/UV Absorption}
During the last decade, largely because of moderate-resolution, 
high-sensitivity UV spectroscopy with \HST\ and soft X-ray spectroscopy with 
\ASCA, it has been recognized that
UV and X-ray absorption features are ubiquitous properties of 
low-luminosity AGNs (e.g., Crenshaw et al.\ 1999).
The origin of the resonance-line and ionization-edge absorption 
features is poorly understood. In many instances, a strong case can be 
made that the UV and X-ray features arise co-spatially. The absorbing 
gas is always blueshifted relative to the emission lines, and multiple 
velocity components are often identifiable in the UV lines. The large 
column densities indicate that
the flows are massive\footnote{This conclusion is dependent on
the assumed ``covering factor'', but this must be of order unity since 
absorption is present in virtually all nearby well-studied AGNs.}, and 
in some cases can involve kinetic energy fluxes similar to the 
radiative output of AGNs. These
properties suggest that they are the analogs of the polar outflows seen
in young stars, i.e., they are a by-product of the accretion process.
They may be somehow related to the much more massive outflows seen
in about 10\% of high-luminosity AGNs, those known as
broad absorption line (BAL) QSOs. Perhaps most importantly,
outflows from AGNs may have a profound feedback effect on star-formation
processes in the host galaxies (e.g., Silk \& Rees 1998).

The basic questions that need to be addressed are:
\begin{enumerate}
\item How much mass and kinetic energy is involved in these outflows, 
and how does this compare to the radiative output of AGNs?
\item How do the properties of the absorbers vary with other AGN 
properties, especially mass, luminosity, and radio loudness?
\end{enumerate}
Only a very large space-based telescope can address these questions, as 
the sources are faint and high spectral resolution observations  
are required. Spectral resolution must be
high enough to resolve the individual velocity components at their 
thermal width ($\sim10$\,\kms, or $R = 30,000$), and the most
important features are the UV resonance lines of abundant elements,
notably \civ, \siIV, \nv, and \ovi. It is only in the lower-luminosity,
lower-redshift objects that the individual velocity components 
are distinguishable from one another; we cannot address the same
question simply by observing similar systems at higher redshift,
as the absorption systems seen in the higher-redshift,
higher-luminosity BALs are virtually continuous in velocity,
making physical analysis extremely difficult and model dependent.

Variability of absorption lines in lower-luminosity AGNs,
which occurs on time scales as short as a day,
and detection and measurement of weak fine structure lines
afford useful probes of physical conditions 
(mainly particle density) in the absorbing gas.
Large collecting apertures are key to acquiring data
of sufficient quality to utilize these tools.

\section{Space Astronomy Infrastructure}
A generally recognized principle in ground-based 
astronomy is that to make efficient use of 
new-technology very large telescopes, we must 
off-load essential work that can be done with smaller 
telescopes. During the early years of \HST, the
{\it International Ultraviolet Explorer (IUE)}
served in this capacity. At the present time,
only \HST\ and the {\em Far Ultraviolet Spectroscopic Explorer
(FUSE)}, which do not have overlapping
capabilities, are operational. By no later than
2010, there is likely to be {\em no} UV spectroscopic
instrument generally available to the community.
Even optimistic launch dates for an 8-m class
UV telescope might mean a long hiatus during which
no UV observations are possible and hard-won expertise
in UV spectroscopy begins to evaporate. Moreover,
once an 8-m UV/optical telescope becomes operational,
either it will spend some fraction of its time doing
truly critical UV observations that might equally well
be done on a smaller, cheaper telescope, or this important
work simply will not get done.

NASA's contribution to astrophysics has been
fantastically successful; in many areas of astronomy, 
and active galactic nuclei constitute only one example,
UV data are simply too critical to do without. 
Large areas of astrophysics opened by UV astronomy
cannot be allowed to die or languish when \HST\
reaches the end of its lifetime. Certainly
the case put forward at this meeting tells us that
there are many frontiers in which we can progress
significantly with an order-of-magnitude increase
in collecting area and a factor of a few improvement
in angular resolution.
However, in our planning, we need to recognize
the need for smaller workhorse facilities as part of the 
space astronomy infrastructure. These do not be
need to be glamorous telescopes ladden with superlatives
or science programs designed to address in a
definitive way one of the handful of current big mysteries
of the Universe. Neither would they be expensive;
indeed, a  1-m class UV spectroscopic telescope fits into a 
Medium Explorer funding envelope ($\sim$\$200M) and expendable
launch vehicle  (e.g., Delta II with a 3-m fairing).
In other words, we are fully capable of solving this problem
essentially with existing resources because new funding lines would not 
necessarily be required. If we fail to solve this problem,
we have no one but ourselves to blame.

\section{Summary}

The most dramatic impact that an 8-m space-based UV/optical
telescope would have on AGNs would be to enable stellar-dynamical
mass measurements out to a distance of $\sim 100$\,Mpc,
a volume large enough to include several AGNs with
reverberation-based mass measurements. The increase in collecting
area relative to \HST\ will enable far more detailed studies
of the poorly understood, but energetically important, massive
outflows seen in AGN spectra. 
The order-of-magnitude
improvement in spatial resolution afforded by \HST\ 
relative to ground-based observations has
provided us with a wealth of information on the inner structure of AGNs
(e.g., Pogge \& Martini 2002) and on the
evolution of AGN host galaxies.
Certainly, another factor of a few improvement in resolution 
and the much larger collecting area will allow us push 
these frontiers forward.

\acknowledgments
I wish to thank my colleagues D.M.\ Crenshaw, L.\ Ferrarese,
F.\ Hamann, L.C.\ Ho, A.\ Laor, P.T.\ O'Brien, P.S.\ Osmer,
R.W.\ Pogge, J.C.\ Shields, and M.\ Vestergaard for helpful discussions 
and suggestions.


\begin{references}

\reference{}Blandford, R., \& McKee, C.F. 1982, ApJ, 255, 419 
\reference{}Crenshaw, D.M., Kraemer, S.B., Boggess, A., Maran, S.P.,
	Mushotzky, R.F., \& Wu, C.-C. 1999, ApJ, 516, 750
\reference{}Ferrarese, L.,  2002, in Current High-Energy
	Emission Around Black Holes, ed.\ C.-H.\ Lee
	(Singapore: World Scientific), in press (astro-ph/0203047)
\reference{}Ferrarese, L., \& Merritt, D. 2000, ApJ, 539, L9
\reference{}Ferrarese, L., Pogge, R.W., Peterson, B.M.,
	Merritt, D., Wandel, A., \& Joseph, C.L. 2001, ApJ, 555, L79
\reference{}Gebhardt, K., et al. 2000a, ApJ, 539, 13
\reference{}Gebhardt, K., et al. 2000b, ApJ, 543, L5
\reference{}Horne, K., Peterson, B.M., Collier, S.J., \& Netzer, H.
	2002, submitted to PASP (astro-ph/0201182)
\reference{}Kormendy, J., \& Richstone, D. 1995, ARAA, 33, 581
\reference{}Malkan, M.A., \& Sargent, W.L.W. 1982, ApJ, 254, 22

\reference{}Miyoshi, M., Moran, J., Herrnstein, J.,
	Greenhill, L, Nakai, N., Diamond, P., \& Inoue, M. 1995, Nature, 
	373, 127	
\reference{}Onken, C.A., \& Peterson, B.M. 2002, ApJ, 572, 746
\reference{}Peterson, B.M. 1993, PASP, 105, 247 
\reference{}Peterson, B.M. 2001, in 
	Advanced Lectures on the Starburst--AGN Connection,
	ed.\ I.\ Aretxaga, D. Kunth, \& R.\ M\'{u}jica,
	(Singapore: World Scientific), p.\ 3
\reference{}Peterson, B.M., \& Wandel, A. 1999, ApJ, 521, L95
\reference{}Peterson, B.M., \& Wandel, A. 2000, ApJ, 540, L13 
\reference{}Pogge, R.W., \& Martini, P. 2002, ApJ, 569, 624
\reference{}Silk, J., \& Rees, M.J. 1998, A\&A, 331, L1
\reference{}Shields, G.A., 1978, Nature, 272, 706
\reference{}Vestergaard, M. 2002, ApJ, 571, 733
\reference{}Wandel, A., Peterson, B.M., \& Malkan, M.A. 1999,
ApJ, 526, 579
\end{references}
\end{document}